\documentclass[conference]{IEEEtran}
\IEEEoverridecommandlockouts
\usepackage{cite}
\usepackage{amsmath,amssymb,amsfonts}
\usepackage{algorithmic}
\usepackage{graphicx}
\usepackage{textcomp}
\usepackage{xcolor}
\def\BibTeX{{\rm B\kern-.05em{\sc i\kern-.025em b}\kern-.08em
    T\kern-.1667em\lower.7ex\hbox{E}\kern-.125emX}}
\begin{document}

\title{Pruning-aware Loss Functions for STOI-Optimized Pruned Recurrent Autoencoders for the Compression of the Stimulation Patterns of Cochlear Implants at Zero Delay
\thanks{This work was supported by the Deutsche
Forschungsgemeinschaft (DFG, German Research Foundation) under Project-ID 381895691 (OS 295/7-3).}
}

\author{\IEEEauthorblockN{1\textsuperscript{st} Reemt Hinrichs}
\IEEEauthorblockA{\textit{Institut für Informationsverarbeitung} \\
\textit{Leibniz University}\\
Hannover, Germany \\
hinrichs@tnt.uni-hannover.de}
\and
\IEEEauthorblockN{2\textsuperscript{nd} Jörn Ostermann}
\IEEEauthorblockA{\textit{Institut für Informationsverarbeitung} \\
\textit{Leibniz University}\\
Hannover, Germany }
}

\maketitle

\begin{abstract}
Cochlear implants (CIs) are surgically implanted hearing devices, which allow to restore a sense of hearing in people suffering from profound hearing loss. 
Wireless streaming of audio from external devices to CI signal processors has become common place. Specialized compression based on the stimulation patterns of a CI by deep recurrent autoencoders can decrease the power consumption in such a wireless streaming application through bit-rate reduction at zero latency.

While previous research achieved considerable bit-rate reductions, model sizes were ignored, which can be of crucial importance in hearing-aids due to their limited computational resources. This work investigates maximizing objective speech intelligibility of the coded stimulation patterns of deep recurrent autoencoders while minimizing model size. For this purpose, a pruning-aware loss is proposed, which captures the impact of pruning during training. 
This training with a pruning-aware loss is compared to conventional magnitude-informed pruning and
is found to yield considerable improvements in objective intelligibility, especially at higher pruning rates.
After fine-tuning, little to no degradation of objective intelligibility is observed up to a pruning rate of about 55\,\%. The proposed pruning-aware loss yields substantial gains in objective speech intelligibility scores after pruning compared to the magnitude-informed baseline for pruning rates above 45\,\%.

\end{abstract}


\section{Introduction}
Cochlear implants (CIs) are surgically implanted hearing-aids capable of restoring a sense of hearing in people suffering from moderate to profound hearing loss. While good speech understanding is achieved in high speech-to-background noise environments, more challenging environments as encountered in social situations still pose a problem \cite{Goehring19}. Wireless streaming of audio as required for, e.g., beamformers, remote microphones \cite{Henry21} or binaural sound coding strategies \cite{Gajecki18} is among the techniques applied to improve speech understanding in these challenging environments. 
 To save power or bandwidth in this wireless transmission, data compression is commonly applied to reduce the bit-rate of  audio signals before transmission. This compression usually introduces an additional delay and should be kept as small as possible, as speech perception of hearing aid users can be affected by  delays above the range of $5-10$ ms  \cite{Stone08}. 
 For this purpose, we proposed \cite{Hinrichs19, Hinrichs21,Hinrichs22a,Hinrichs22b} to code and transmit the electrical stimulation patterns generated by the sound coding strategy of the CI. With this approach, we achieved a bitrate of about 4.7 kbit/s at zero latency with little to no  degradation in objective intelligibility scores \cite{ Hinrichs23}.

The commercial application of deep learning in hearing aids remains limited, despite a vast amount of research being published in this area, in part due to limited computational resources. For example, hearing aid processors offer tiny amounts of on-chip memory of around 32 kB to 300 kB \cite{RAMHearingAid}.
This complicates the usage of deep neural networks, which consists of  thousands up to billions of parameters, depending on model architecture, and as such usually require substantial  amounts of memory and further computational resources. 
To tackle such limitations, compression of artificial neural networks, namely by quantization and/or pruning, is commonly employed to allow the deployment on resource-constraint devices.

Pruning methods are roughly  differentiated according to pruning structure (structured vs. unstructured), pruning time (before, during or after training) and pruning criteria. A recent survey of pruning methods can be found in \cite{Cheng24}. 

Pruning criteria are split between learned and non-learned methods. Learned pruning methods consist of, e.g., reinforcement learning for layerwise sparsity ratio selection followed by conventional pruning \cite{He2018AMCAF}, learning meta-networks pruned using evolutionary algorithms \cite{Liu2019MetaPruningML} and sparsity-regularization methods that introduce additional sparsity-related terms into the loss function \cite{Dery24, He17,PruningHinrichs22}.
Non-learned pruning methods are largely based on assessing the importance of weights through derivatives of the loss function.  The  idea is to prune the weights with the least impact on the loss function.
The most basic approach  of this class of pruning methods uses magnitude-informed pruning, where weights with the smallest absolute value are pruned.
Further, commonly employed pruning methods are gradient-informed, where weights are pruned based on the absolute gradients, and so called movement pruning, where weights are pruned based on the magnitude of the product the weights and their respective gradients $|\omega_i \cdot \frac{\partial \mathcal{L}}{\partial \omega_i}|$.

More generally, such methods  are built on the belief that Taylor's expansion of a loss function allows to reasonably assess the loss change due to pruning. 
However, derivatives, evaluated at a single point, give local information only.

\subsection{Contribution}
The previously mentioned class of pruning methods, based on derivatives of the loss function, has two major drawbacks: Besides being unable to capture the global loss change due to pruning weights, these methods usually do not allow the network to adapt to the weights to be pruned during the initial training.
If the network "knew" it was going to be pruned, it could adjust its weights to become more robust to the pruning of its weights.
These two drawbacks can be improved in the following way:
By introducing weight perturbation during training, the global impact of subsequent pruning of certain weights can be captured during training.
As such, the network can reconfigure its weights to become more robust to said pruning.

The proposed method is to devise a pruning-aware loss, which incorporates the above ideas. The proposed method makes no assumptions regarding model architecture (e.g. does not differentiate between layer types) and requires only a moderately increased training complexity.

\section{Methods and Materials}

\subsection{Pruning}
Pruning in the context of artificial neural networks is any mapping
\begin{equation}
    P:\omega \rightarrow \hat{\omega}
\end{equation}
of a network's weights $\omega\in \mathbb{R}^N$, with the number of weights $N$, to the pruned weights $\hat{\omega}\in \mathbb{R}^N$, where the mapping $P$ can be defined according to
\begin{equation}
    P(\omega)_i \equiv \hat{\omega}_i := \begin{cases}
    0   &  i\in I_{pruned}\\
        \omega_i & \text{otherwise}.
    \end{cases}
\end{equation}
$I_{pruned}\subset \{1,2,\dots, N\}$ is a set of indices indicating the network weights to be pruned/set to zero. Weights not pruned remain unchanged.
A pruned network as such is a neural network, where the pruned weights are removed, potentially reducing the complexity, but in general also the performance, of a neural network.

Pruning methods usually consist of the initial, actual pruning as defined above, and subsequent fine-tuning of the pruned network. 
The above pruning mapping $P$ can, without loss of generality, be written in the form
\begin{equation}
    P(\omega) = \omega + \Delta\omega,
\end{equation}
with 
\begin{equation}
    \Delta\omega_i = \begin{cases}
        -\omega_i & i\in I_{pruned}\\
        0 & \text{otherwise}
    \end{cases}.
\end{equation}
A key component of pruning methods therefore usually is to find an optimal $\Delta\omega$, which  hereinafter is called the pruning direction.

% A common class of pruning methods bases the choice of the pruning direction on derivatives of the loss function.  The  idea is to prune the weights with the least impact on the loss function.
% The most basic approach  of this class of pruning methods uses magnitude-informed pruning, where the pruning direction is chosen, such that the weights with the smallest absolute value are pruned.
% Further, commonly employed pruning methods are gradient-informed, where weights are pruned based on the absolute gradients , and so called movement pruning, where weights are pruned based on the magnitude of the product the weights and their respective gradients $|\omega_i \cdot \frac{\partial \mathcal{L}}{\partial \omega_i}|$.

% More generally such methods  are built on the belief that Taylor's expansion of a loss function allows to reasonably assess the loss change due to pruning. 
% However, derivatives, evaluated at a single point, give local information only. As such, pruning directions based on derivatives can be misguided, as the derivatives can change arbitrarily much for minor deviations from the point of evaluation. 

\subsubsection{Proposed Pruning Method}
In this work, we propose a novel pruning method based on the observation that one issue of pruning methods is the choice of an optimal pruning direction. 

% The previously mentioned class of pruning methods, based on derivatives of the loss function, has two major drawbacks: besides being unable to capture the global loss change due to pruning weights, these methods usually do not allow the network to adapt to the pruning direction during training.
% If the network "knew" it was going to be pruned in some given direction, it could adjust its weights to become more robust to pruning in this direction.

% These two drawbacks can be improved on in the following way:
% By introducing weight perturbation during training, the global impact of subsequent pruning of certain weights can be captured during training.
% As such, the network can reconfigure its weights to become more robust to said pruning.

It is proposed  to devise a pruning-aware loss, which incorporates the above ideas.
To make a given loss $\mathcal{L}_\omega$ pruning-aware (PA), where $\omega$ denotes a network's weights, the loss function is modified according to

\begin{equation}
\label{eq:pruningawareloss}
    \mathcal{L}_{\omega_n}^{PA} = \mathcal{L}_{\omega_n} + \lambda  |\mathcal{L}_{\omega_n} - \mathcal{L}_{\omega_n + \Delta \omega_n}|.
\end{equation}

$\mathcal{L}_{\omega_n + \Delta \omega_n}$ is the loss of a network  computed with the perturbed weights $\omega_n + \Delta \omega_n$, where $n$ is the iteration index. $\lambda$ is a positive weighting factor, in this work equal to one. The term $|\mathcal{L}_{\omega_n} - \mathcal{L}_{\omega_n + \Delta \omega_n}|$ captures the global impact of pruning. 

Because it appears reasonable to assume,that strong perturbation of the weights during the early phases of training could not give a network enough time to reconfigure itself, pruning is gradually introduced in the loss. 
That is, the pruning direction $\Delta \omega_n$ is gradually approaching the actual pruning direction across training iterations according to 
\begin{equation}
\label{eq:perturbation}
    \Delta\omega_n = g(\frac{n}{n_{max}})  \Delta\omega_{magn,n},
\end{equation}

where $\Delta\omega_{magn,n}$ is the magnitude-informed pruning direction in iteration $n$, i.e., the weights with the smallest magnitude in iteration $n$ are perturbed. $g:[0,1]\rightarrow [0,1]$ is called perturbation function and defines the attack time of the weight perturbation.
$n_{max}$ is the total number of training iterations.

Because the perturbation is applied to all weights, indifferent of the layers they belong to (global unstructured pruning), the proposed method is presumed model agnostic, i.e., can be applied to any model architecture. A downside seems to be the moderately increased complexity due to either additional training steps or the need of an additional gradient computation.

% Training a model for additional epochs with this pruning-aware loss should automatically yield a pruning-robust network, which can then be pruned using the magnitude-informed pruning direction.

To the best knowledge of the authors, no similar pruning method exists in the literature. 

\subsection{Evaluation Metric}
Objective speech intelligibility of the decoded stimulation patterns is assessed using the vocoder short-time objective intelligibility measure (VSTOI). For its application, stimulation patterns are resynthesized by a (sine) vocoder provided by the Nucleus Matlab Toolbox \cite{NMT}. The resulting waveform is then compared to a corresponding clean speech signal using the well-known short-time objective intelligibility measure (STOI) \cite{STOI}. STOI returns scores in the range $[0,1]$ with $0$ indicating worst, and $1$ indicating best intelligibility.

VSTOI is known to be very sensitive, that is, minor changes in VSTOI scores correspond to way larger changes in word recognition scores as tested in subjects. Going by word-recognition functions shown in \cite{Watkins18}, in the linear region, one can deduce, that  a change of approximately 0.006 in VSTOI score corresponds to about 5\,\% in  word recognition scores. Note, that the precise correspondence between VSTOI  and word recognition scores generally  depends on the dataset being used.

\subsection{Stochastic Perturbation Simultaneous Approximation}
\label{ssec:NA_STOI}
For all optimizations in this work, the Stochastic Perturbation Simultaneous Approximation (SPSA) algorithm is used \cite{SPSA}. The SPSA allows to optimize non-differentiable functions through random perturbations and iterative updates.

 The update equation of the SPSA for all parameters $\underline{\omega}$ of a model (including, e.g., quantizers) is
\begin{equation}
\label{eq:NA_update}
    \underline{\omega}_{k+1} = \underline{\omega}_k + a_k \frac{(y_{k}^+ - y_{k}^-)}{2\cdot c_k}\Delta_k,
\end{equation}
where $y_{k}^{\pm} = f(\underline{\omega}_k \pm c_k\Delta_k)$, $\Delta_k\in\{-1,1\}^N$ a vector of iid noise, $a_k,c_k>0$ with $a_k,c_k\rightarrow 0$. N is the total number of parameters. 
In our work, we used $a_k = \frac{a}{(A+k+1)^\gamma} $ with $a=1$, $A=10273$ and $\gamma = 0.602$ as well as $c_k = \frac{c}{(k+1)^\beta}$ with  $\beta = 0.101$ and $c=0.020765$. $A$ and $c$ were found in \cite{Hinrichs23} through hyperparameter optimization.
In our case, the function $f$ returns mean VSTOI scores of the decoded stimulation patterns of the model to be optimized.

\subsection{Evaluation}
To assess the impact of the proposed pruning-aware training, a pretrained feedback recurrent autoencoder (FRAE) with 6 bit vector quantization as described in \cite{Hinrichs23} is further optimized using the proposed pruning-aware loss according to Eq. \ref{eq:pruningawareloss} for 1000 iterations.  
The following perturbation functions were assessed: $g(x) = x$, $g(x)=x^2$ and $g(x)=x^3$. This choice yields more and less "aggressive" perturbation of the network weights.

We investigate whole-model pruning and decoder-only pruning.
In whole-model pruning, the weights of both, encoder and decoder, are eligible for pruning. In decoder-only pruning, only the weights of the decoder are eligible, the weights of the encoder remain unchanged.
This distinction is motivated by the observation that a CI signal processor, acting solely as a receiver as in wireless phone call streaming,  only needs to decode the received data, thus only needs the FRAE decoder. 
In bilateral signal processing strategies, however, where stimulation patterns could be transmitted between two signal processors located at opposite ears, the signal processors have to encode and decode data, as such require the entire FRAE model. 

Pruning rates $pr\in\{0.05, 0.1, \dots, 0.95\}$ are investigated. Only pruning of the weights (in contrast to bias pruning) is considered.
For decoder-only pruning, the pruning rate refers solely to the decoder's weights. 

The FRAE model used in this work is an already tiny model with only slightly more than 3300 weights. 
Due to this, unlike for way larger models like ResNets,
it cannot be expected to achieve little to no degredation in model performance for very high pruning rates like 95\,\%.

As baseline for our method we chose magnitude-informed pruning.

The entire evaluation consist of pruning and subsequent fine-tuning of the pruned models. 
For the proposed method, this procedure consists of an initial 1000 SPSA iterations, where the pruning-aware loss as given in Eq. \ref{eq:pruningawareloss} is used, with the mean VSTOI scores of the decoded stimulation patters of the training set as base loss function $\mathcal{L}_{\omega}$. After this training phase, the FRAE is pruned with a predefined pruning rate, and the pruned FRAE is trained, now with respect to mean VSTOI scores without any perturbation, for another 7000 iterations.

For the  baseline, the FRAE is pruned first, without any additional training, and subsequently fine-tuned for 8000 iterations using the SPSA algorithm to match the total of 8000 iterations of the proposed method.

\subsection{Dataset}
\label{ssec:ACE}
The dataset of this work is identical to the one used in \cite{Hinrichs23}. It consists of stimulation patterns derived from speech in noise samples, where the speech samples stem from the well-known TIMIT speech corpus. The speech samples were mixed with noise at signal-to-noise ratios ranging from -5 dB to 40 dB using head-related transfer functions. Office, bus, restaurant and CCITT-noise were considered. 

The stimulation patterns are derived from these audio mixes using the advanced combination encoder (ACE) sound coding strategy as provided by the  Nucleus Matlab Toolbox. Default parameters are used for ACE, most importantly, a channel stimulation rate of 900 pulses per second as well as $N=8$ and $M=22$ is used, where $M$ is the total number of subbands and $N$ is the maximum number of selected subbands per frame. The dataset and ACE are described with considerable detail in \cite{Hinrichs23} and \cite{Hinichs24}.  

The resulting train set consists of 100 stimulation patterns, a sufficiently generalizing subset of a more than 4000 files counting training data set, while the test set consists of 1680 stimulation patterns. All presented results are obtained on the test set.

\begin{figure}
    \centering
    \includegraphics[width=\linewidth]{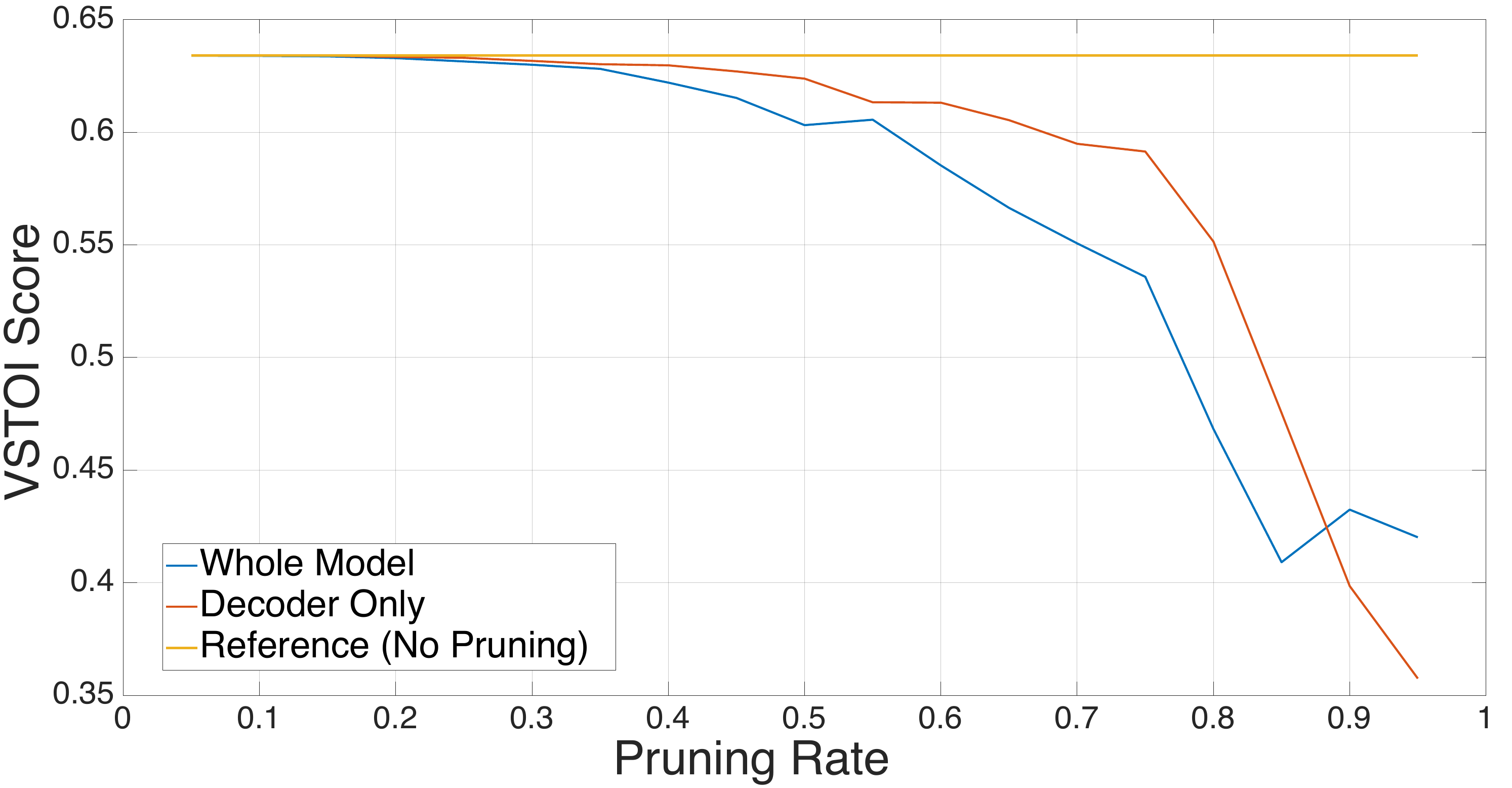}
    \caption{Baseline magnitude-informed prunings results for whole-model and decoder-only pruning.}
    \label{fig:baseline}
\end{figure}

\section{Results}
Baseline results are shown in Fig. \ref{fig:baseline} for whole-model and decoder-only pruning alongside the reference model performance, i.e., the original model without any applied pruning. As expected, decoder-only pruning yields considerably higher post-pruning VSTOI scores. However, until a pruning rate of about 35\,\%, pruning barely affects model performance for either pruning type. At  pruning rates of 90\,\% and 95\,\%, the performance of whole-model pruning surpasses decoder-only pruning with respect to VSTOI scores. However, VSTOI scores below about 0.45 are achieved by randomly initialized models. As such, the decoded signals contain only noise and do not represent a meaningful compression scheme.
\begin{figure}
    \centering
    \includegraphics[width=\linewidth]{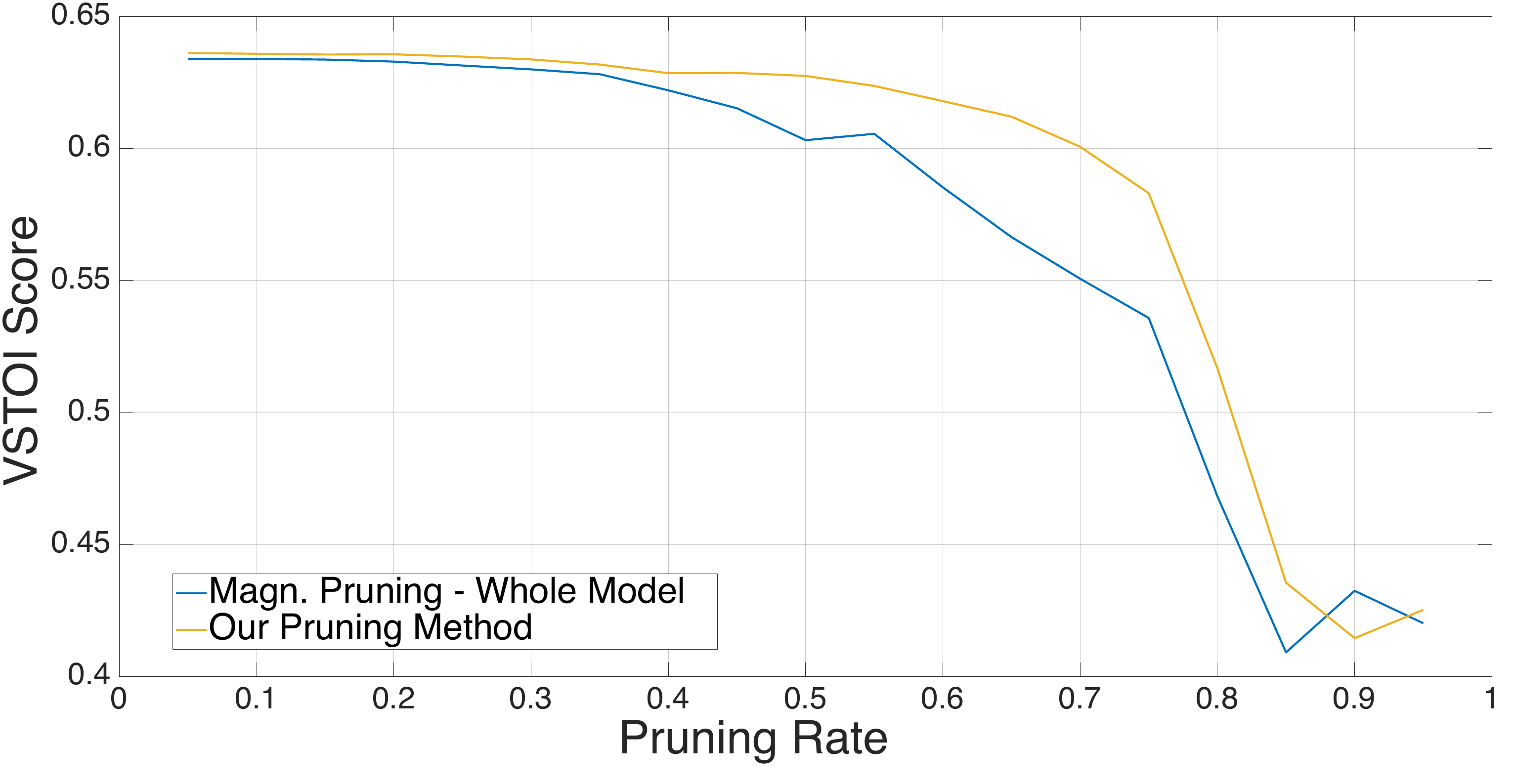}
    \caption{VSTOI scores of the baseline whole-model pruning and the proposed pruning method. Similar performance is achieved up to a pruning rate of 35\,\%. Starting at 40\,\%, with a single exception, the proposed pruning method considerably outperforms the baseline method.  }
    \label{fig:pruningCompareWholemodel}
\end{figure}
\begin{figure}[b]
    \centering
    \includegraphics[width=\linewidth]{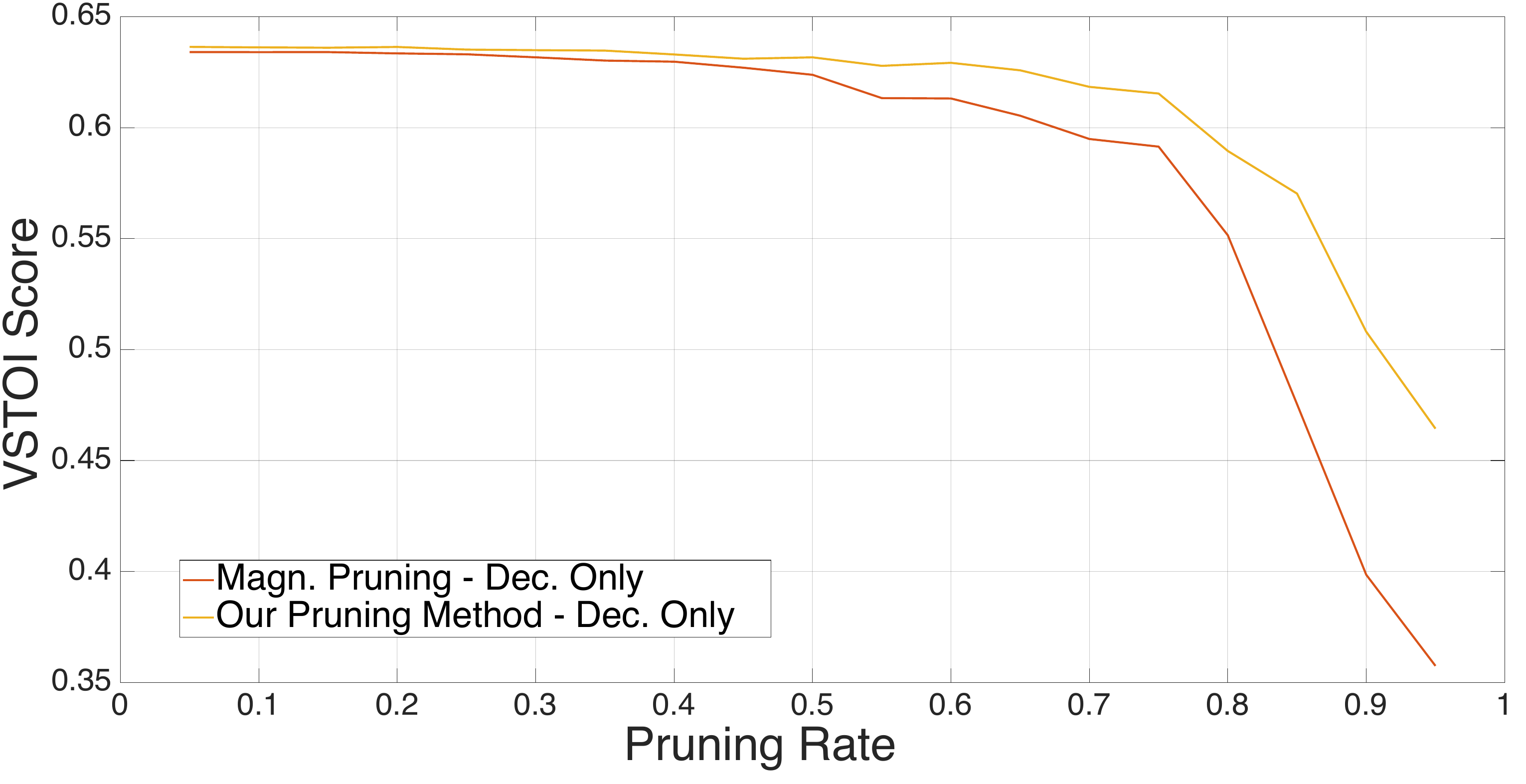}
    \caption{VSTOI scores of the baseline decoder-only pruning and the proposed pruning method. Similar performance is achieved up to a pruning rate of 45\,\%. Starting at 50\,\%,  the proposed pruning method considerably outperforms the baseline method.  }
    \label{fig:pruningCompareDecOnly}
\end{figure}
Fig. \ref{fig:pruningCompareWholemodel} shows a comparison of the VSTOI scores of the baseline pruning method and the proposed pruning method for whole-model pruning. If not noted otherwise, linear perturbation was used in the pruning-aware loss in all presented figures. The proposed method yields superior post-pruning performance for all pruning rates with the exception of the pruning rates  90\,\% and 95\,\%, where again the pruning impact is so grave, that the model outputs noise-like signals and the difference in VSTOI scores is meaningless. The benefit of the proposed pruning methods is the largest for a pruning rate of 70\,\%.
\begin{figure}[t]
    \centering
    \includegraphics[width=\linewidth]{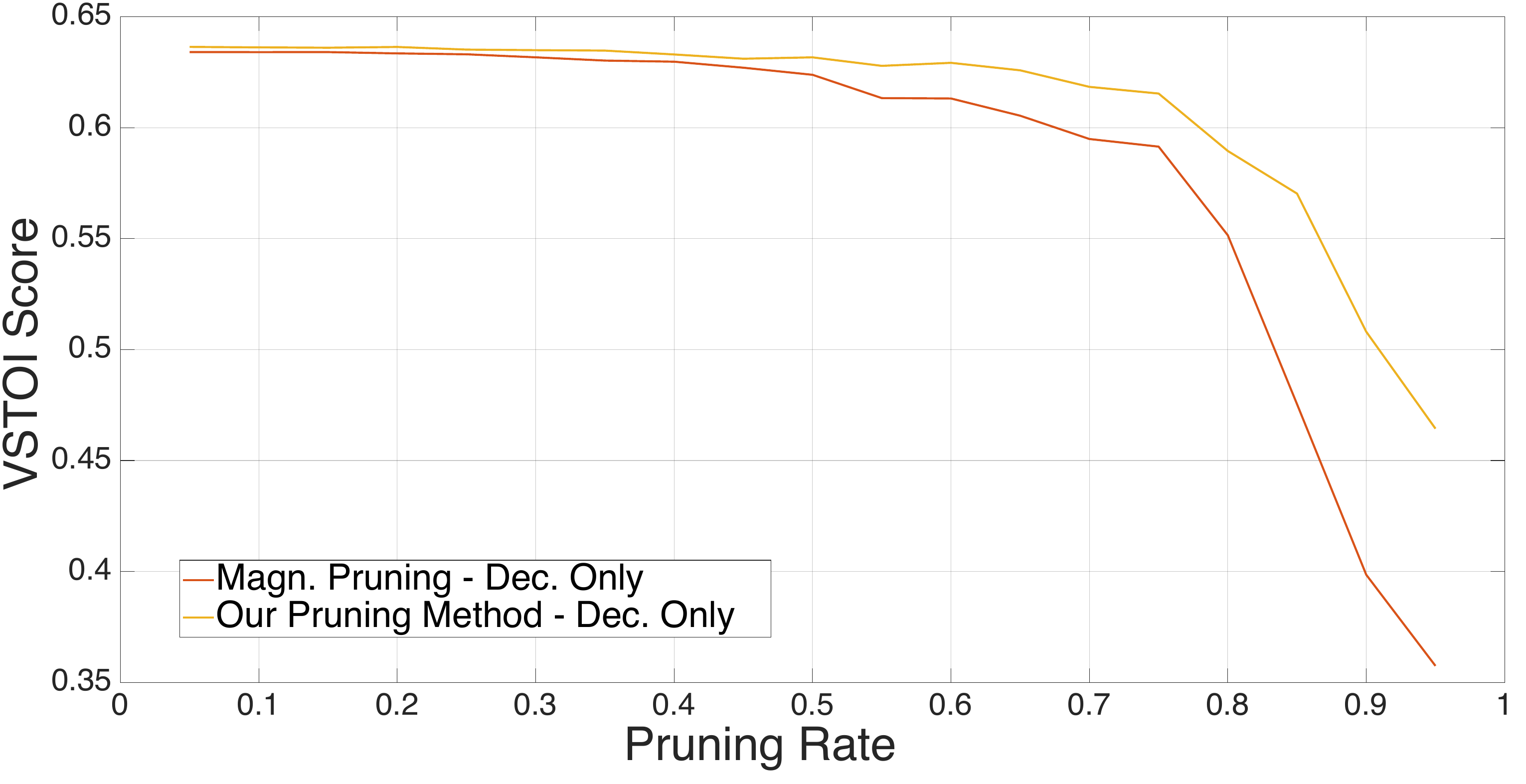}
    \caption{VSTOI scores of the baseline decoder-only pruning and the proposed pruning method. Similar performance is achieved up to a pruning rate of 45\,\%. Starting at 50\,\%,  the proposed pruning method considerably outperforms the baseline method.  }
    \label{fig:pruningCompareDecOnly}
\end{figure}
Fig. \ref{fig:pruningCompareDecOnly} compares post-pruning VSTOI scores across pruning rates for the decoder-only pruning. Again, the proposed pruning method proves considerably superior with little degredation until a pruning rate of 60\,\%, compared to about 45\,\% for the baseline method.   
\begin{figure}[b]
    \centering
    \includegraphics[width=\linewidth]{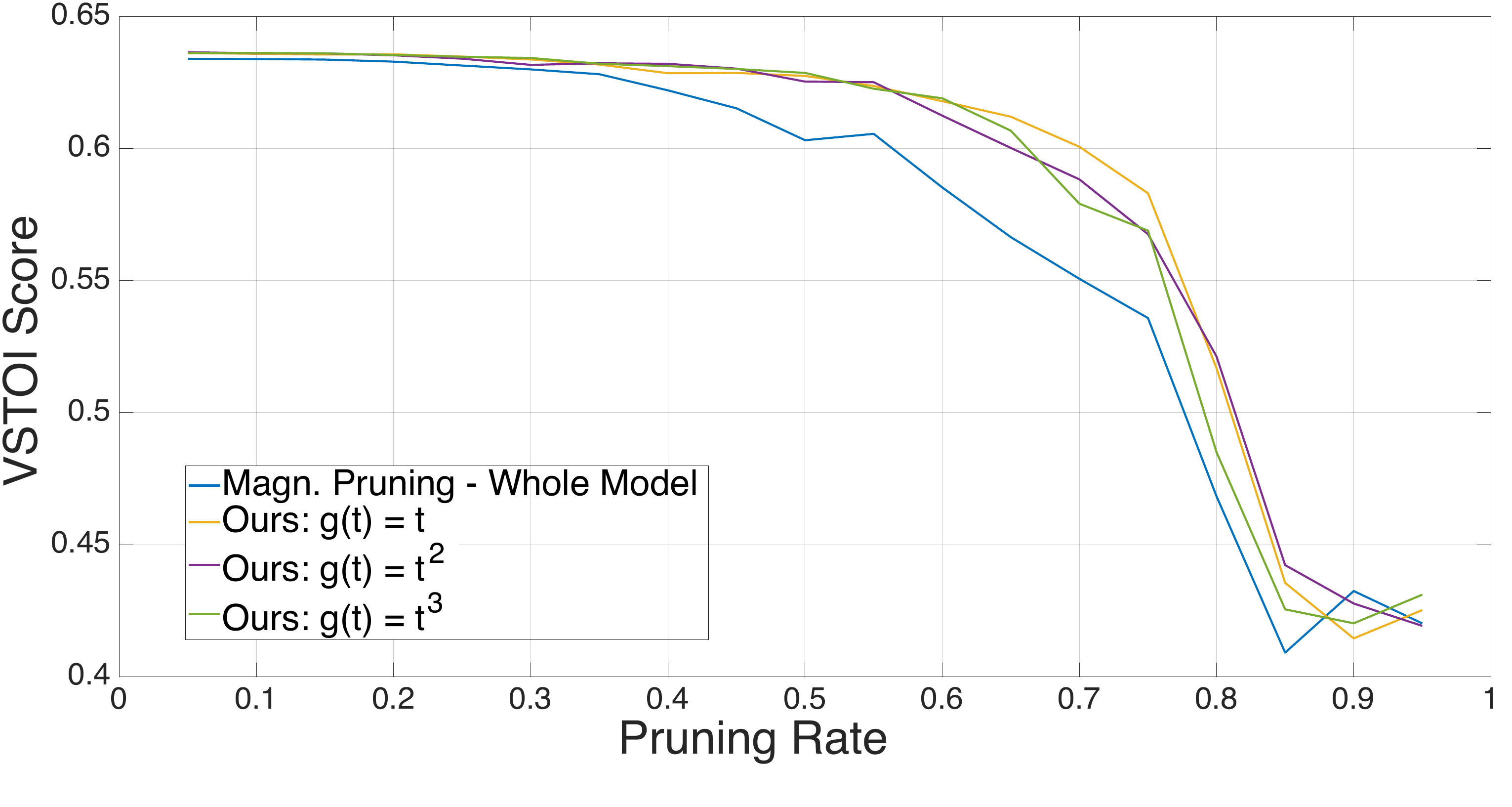}
    \caption{VSTOI scores for baseline magnitude-informed pruning as well as the proposed pruning methods for the three investigated perturbation functions. While all three functions achieve qualitatively similar results, linear perturbation proves superior for pruning rates between 60\,\% and 80\,\%.}
    \label{fig:perturbationComparison}
\end{figure}
Fig. \ref{fig:perturbationComparison} shows the post-pruning VSTOI scores for the three investigated perturbation functions as well as the baseline method for the whole-model case.
Qualitatively, all methods show identical performance, the baseline method is outperformed by a considerable margin for either perturbation function. However, there appears to be a slight edge for the linear perturbation function, especially for the pruning rates from 65\,\% to 75\,\%. Results are qualitatively identical for the decoder-only case.

\begin{figure}
    \centering
    \includegraphics[width=\linewidth]{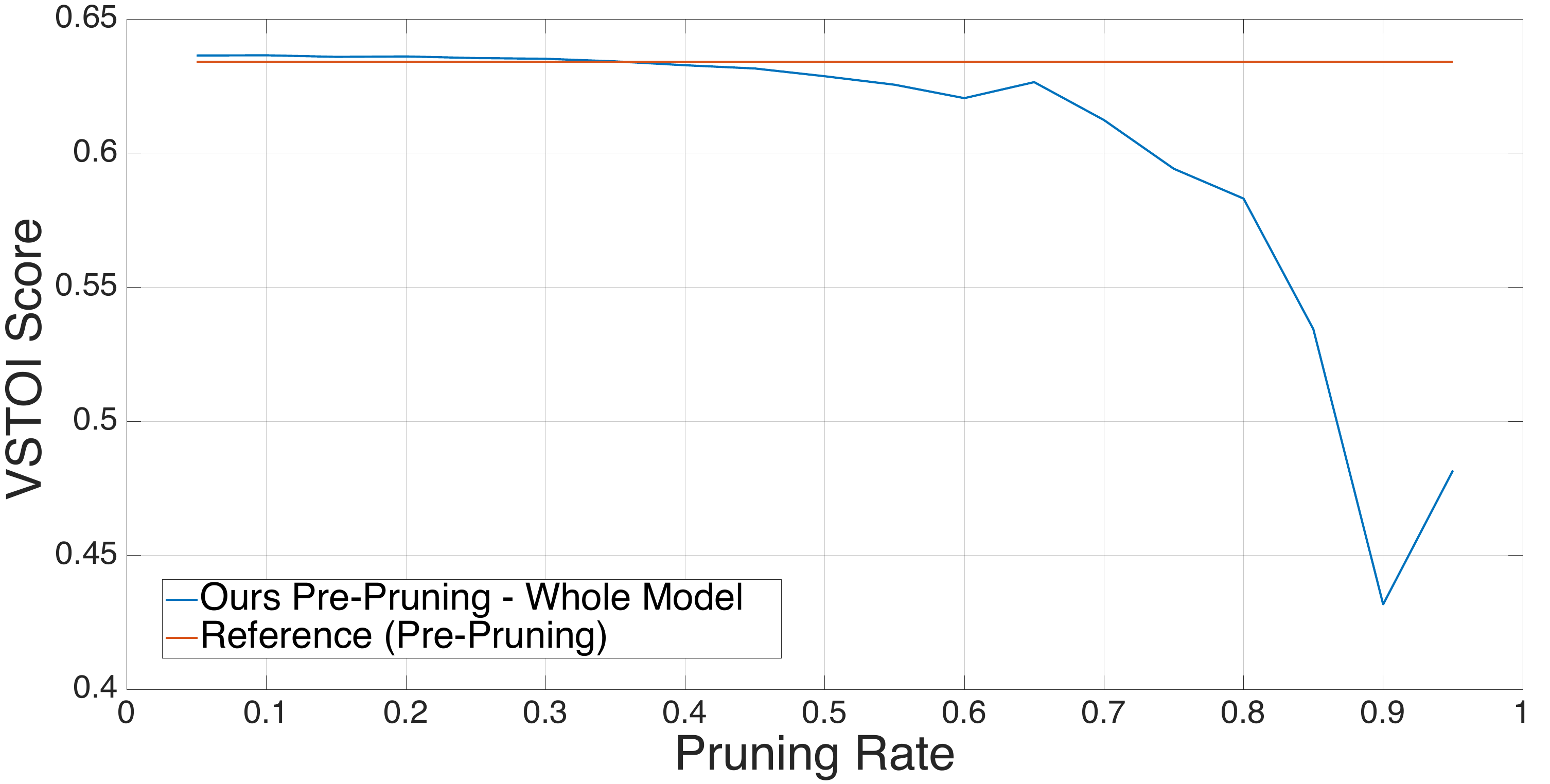}
    \caption{VSTOI scores after training, but before pruning, with the proposed pruning-aware loss (whole-model). As reference, the performance of the original model without pruning is included. After training, VSTOI scores drop for pruning rates above 40\,\%, suggesting that the increased robustness to pruning is not due to improved overall performance. }
    \label{fig:postKWTraining}
\end{figure}

The presented advantage of the proposed pruning method is not due to training for another 1000 epochs. Fig. \ref{fig:postKWTraining} shows VSTOI scores after training but before pruning for the proposed pruning method and the whole-model pruning case. The Figure includes the VSTOI score of the original, unpruned model. 
Note, that the desired pruning rate affects the training through Eq. \ref{eq:perturbation}.
While model performance improved slightly with respect to VSTOI scores for pruning rates below 40\,\%, model performance  actually decreased due to the additional training. 
This shows that the model indeed becomes more robust to pruning due to the proposed additional training step, because, despite lower VSTOI scores before pruning, post-pruning scores are superior for the proposed method.

\begin{figure}[b]
    \centering
    \includegraphics[width=\linewidth]{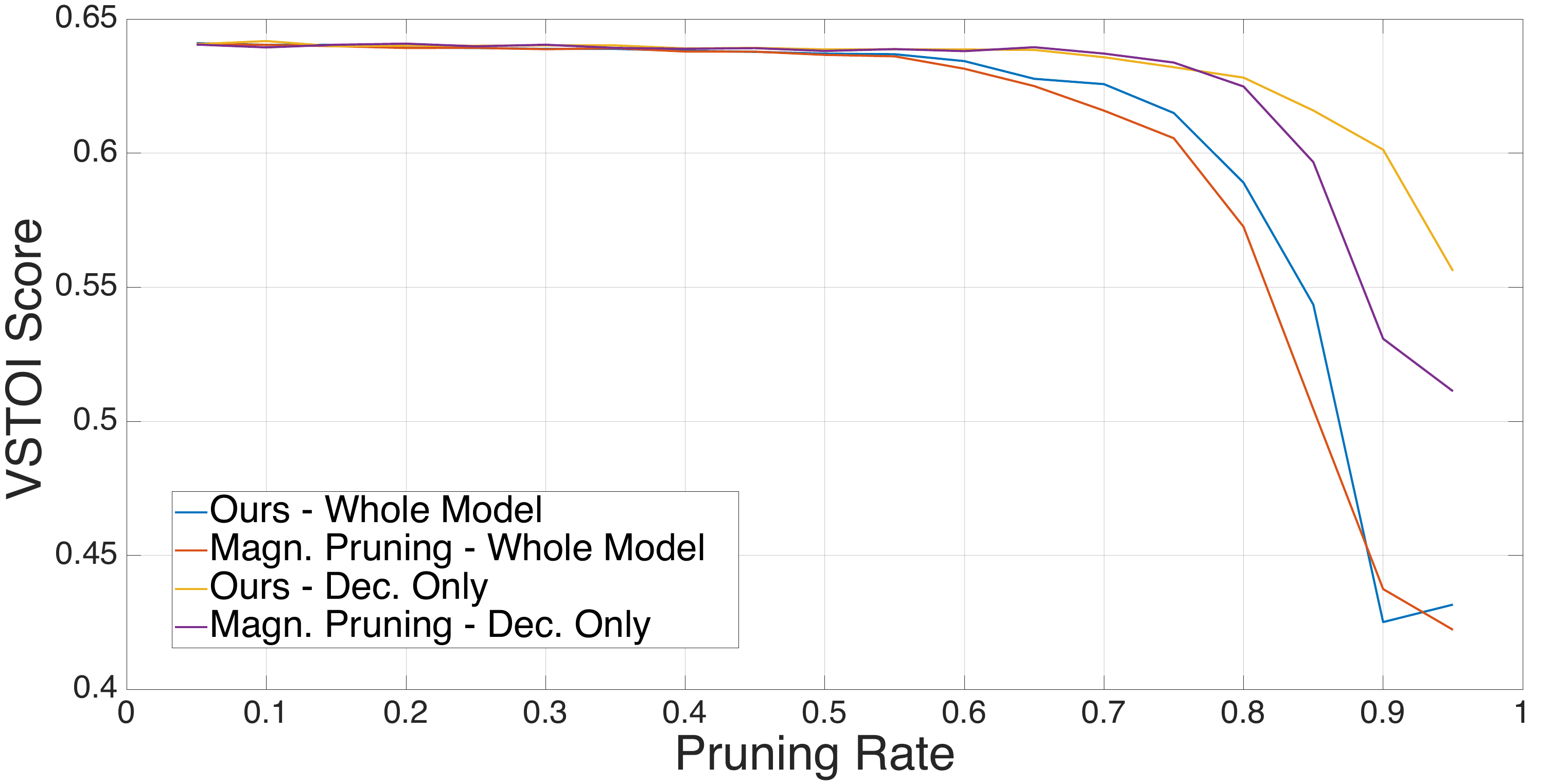}
    \caption{VSTOI scores after pruning and subsequent fine-tuning for whole-model and decoder-only pruning. }
    \label{fig:finetuning}
\end{figure}

Finally, Fig. \ref{fig:finetuning} shows results after fine-tuning for the proposed pruning method as well as the baseline pruning method for the whole-model and decoder-only case. 
The performance gap closes considerably, but remains substantial for whole-model pruning between a pruning rate of 70\,\% to 85\,\%. The largest difference in VSTOI scores of 0.039 is achieved at a pruning rate of 85\,\%.
As before, model performance  at a pruning rate of 90\,\% and 95\,\% is so poor that the observed differences can not be used to reasonably compare the investigated methods.

For decoder-only pruning, up to a pruning rate of 80\,\%, VSTOI scores are almost identical. Surprisingly, despite the gap before fine-tuning, for pruning rates between 65\,\% and 75\,\% (including), the baseline method achieves slightly superior VSTOI scores with a  difference of at most 0.0017 for a pruning rate of 75\,\%. Such a difference is not significant regarding speech intelligibility and could likely arise due to chance as the  SPSA algorithm, due to its random perturbations, can not guarantee identical optimization results.

\section{Discussion}
%This work proposes and investigates a pruning-aware loss for neural network pruning. Training a network with such a pruning-aware loss (and subsequent magnitude-informed pruning) is compared to magnitude-informed pruning without specialized training. 

The impact of the pruning-aware loss is quite considerable: 
Comparing Fig. \ref{fig:postKWTraining}, which shows the VSTOI scores pre-pruning, but after training with the pruning-aware loss, to Fig. \ref{fig:perturbationComparison}, reveals that -- as hoped for -- the pruning itself has decreasing impact on model performance. The largest decrease in model performance stems from the model decreasing its own performance to achieve pruning robustness. The loss in VSTOI score due to pruning is limited to -0.016 up to a pruning rate of 75\,\%. Even then, pruning reduces model performance by at most -0.085 at a pruning rate of 85\,\% (whole-model), compared to -0.23 for the baseline method. Any additional performance decrease of the proposed pruning method is due to the training step prior to pruning. 

The choice of the perturbation function has a considerable impact and the perturbation function yielding most aggressive weight perturbation during training allowed to achieve the best results.
Accordingly, it is likely that even more aggressive perturbations could allow to achieve even better results. This 
was partially -- but not for all pruning rates -- observed for a root-square function.
Parametrizing the perturbation function though, e.g., monotonic interpolation, could allow to yield optimal pruning-aware loss function.

The benefit of the pruning-aware loss appears to increase with increasing pruning rate until the impact of pruning becomes so grave that the model cannot approximate a reasonable compression algorithm anymore. Therefore, the proposed method should be investigated using larger networks to fully evaluate its potential.

Using the magnitude-informed pruning direction $\Delta\omega$ in Eq. \ref{eq:perturbation} appears to be a natural choice, especially due to the simplicity of its calculation. While it is conceivable to use other pruning-directions, like gradient-informed pruning directions, because the network reconfigures itself to be robust to the chosen direction, there seems to be no obvious reason, why a certain direction should be better than another.

The publication most similar to our work appears to be a previous work of our own \cite{PruningHinrichs22}. There, a hessian-related term was introduced into the loss function to train a network to be robust towards weight quantization (and perturbations in general). The method proved to be very effective for quantization and to a lesser degree for pruning. However, in contrast to the proposed method of this work, the computational complexity of the method presented in \cite{PruningHinrichs22} is significantly larger, due to the need of computing the hessian of the loss function. Therefore, for pruning, the proposed pruning-aware loss method of this work is considerable superior.

% \section{Future Work}
% As the linear perturbation function proved to be slightly superior to the other two tested perturbation functions, more aggressive perturbation functions like a square-root may prove beneficial.
% Furthermore, we only investigated simple perturbation functions in this work. It might be an interesting endeavour to parametrize the perturbation function $f$ in Eq. \ref{eq:perturbation} to investigate optimal aggressiveness. Bernstein polynomials or monotone cubic interpolation may be suitable for this task.

\section{Conclusion}
This work proposed and investigated a novel  pruning method for the pruning of artificial neural networks in the context of the compression of the stimulation patterns of cochlear implants. A pruning-aware loss is proposed to automatically achieve pruning-robust networks after training. The proposed method does make no assumptions regarding model architecture.

The evaluation showed improvements in post-pruning performance due to the proposed pruning method, with little degradation in vocoder short-time objective intelligibility scores (VSTOI scores) up to pruning rates of about 40\,\% for whole-model pruning and about 65\,\% for decoder-only pruning. The proposed pruning-aware loss yields substantial gains in VSTOI scores after pruning compared to the magnitude-informed baseline for pruning rates above 45\,\%.

This gap in performance mostly remained, but decreased considerably after fine-tuning. Nevertheless, the proposed method achieved considerably higher VSTOI scores for pruning rates exceeding pruning rates of 70\,\% (whole-model) and 80\,\% (decoder-only).

\bibliographystyle{IEEEtran}

\end{document}